\documentclass[aps,prb,preprint,longbibliography,superscriptaddress]{revtex4-2}
\usepackage{mathrsfs}
\usepackage{amsmath,gensymb}
\usepackage{amsfonts}
\usepackage{amssymb}
\usepackage{amsthm}
\usepackage{graphicx}
\usepackage{natbib}
\usepackage{xcolor}
\usepackage{hyperref}
\usepackage{bm}
\usepackage[caption=false]{subfig}
\usepackage{verbatim}
\usepackage[normalem]{ulem}

\begin{document}

\title{Superconducting spin valves based on antiferromagnet/superconductor/antiferromagnet heterostructures}

\author{G. A. Bobkov}
\affiliation{Moscow Institute of Physics and Technology, Dolgoprudny, 141700 Moscow region, Russia}

\author{V. M. Gordeeva}
\affiliation{Moscow Institute of Physics and Technology, Dolgoprudny, 141700 Moscow region, Russia}

\author{Lina Johnsen Kamra}
\affiliation{Condensed Matter Physics Center (IFIMAC) and Departamento de F\'{i}sica Te\'{o}rica de la Materia Condensada, Universidad Aut\'{o}noma de Madrid, E-28049 Madrid, Spain}
\affiliation{Department of Physics, Massachusetts Institute of Technology, Cambridge, MA-02139,USA}

\author{Simran Chourasia}
\affiliation{Condensed Matter Physics Center (IFIMAC) and Departamento de F\'{i}sica Te\'{o}rica de la Materia Condensada, Universidad Aut\'{o}noma de Madrid, E-28049 Madrid, Spain}

\author{A. M. Bobkov}
\affiliation{Moscow Institute of Physics and Technology, Dolgoprudny, 141700 Moscow region, Russia}

\author{Akashdeep Kamra}
\affiliation{Condensed Matter Physics Center (IFIMAC) and Departamento de F\'{i}sica Te\'{o}rica de la Materia Condensada, Universidad Aut\'{o}noma de Madrid, E-28049 Madrid, Spain}

\author{I. V. Bobkova}
\affiliation{Moscow Institute of Physics and Technology, Dolgoprudny, 141700 Moscow region, Russia}
\affiliation{National Research University Higher School of Economics, 101000 Moscow, Russia}

\begin{abstract}
Proximity effect at superconductor/antiferromagnet (S/AF) interfaces, which manifests itself as generation of N\'eel-type triplet correlations, leads to sensitivity of the superconducting critical temperature to the mutual orientation of the AF N\'eel vectors in AF/S/AF trilayers, which is called the spin-valve effect. Here we predict that the spin-valve effect in AF/S/AF heterostructures crucially depends on the value of the chemical potential of the superconducting interlayer due to the occurrence of the finite-momentum N\'eel triplet correlations. In addition we investigate equal-spin triplet correlations, which appear in AF/S/AF structures for non-aligned N\'eel vectors of the AFs, and their role in the nonmonotonic dependence of the superconducting critical temperature of the AF/S/AF structure on the mutual orientation of the AF N\'eel vectors. The influence of impurities on the spin-valve effect is also investigated.

\end{abstract}

\maketitle

\section{Introduction}
\label{intr}
Heterostructures constructed of superconductors and magnetic materials are objects of great interest for superconducting spintronics due to proximity effects occuring in the nanoscale interface regions \cite{Buzdin2005,Bergeret2005,Eschrig2015,Linder2015}. In particular, magnetic-induced spin-splitting field leads to partial conversion of singlet pairing correlations to triplet ones, which suppresses conventional spin-singlet superconductivity. One of the possible spintronics devices based on such structures is a superconducting spin valve, where the superconducting critical temperature $T_c$ is sensitive to the mutual orientation of the magnetic layers magnetizations (spin-valve effect). In particular, a switching between the superconducting and the normal states, that is the absolute spin-valve effect, can be realized by changing the mutual orientation of the magnetic layers.

Spin valves based on superconductor/ferromagnet (S/F) proximity effect have been widely studied both theoretically and experimentally. The F/S/F structure with insulating ferromagnets was theoretically considered by de Gennes \cite{DEGENNES1966}, who showed that the average exchange field felt by the superconductor is proportional to $\cos{(\phi/2)}$, where $\phi$ is the misorientation angle between F magnetizations. This result predicted the spin-valve effect in such systems: according to it, the critical temperature in the case of parallel (P) magnetizations should be lower than in the antiparallel (AP) configuration, which was later experimentally obtained by Li \textit{et al.} \cite{Li2013}. 

In trilayers with metallic ferromagnets the physics is more complicated because the singlet and triplet superconducting correlations can also penetrate into the ferromagnetic regions due to the proximity effect. However, such structures (F1/F2/S as well as F/S/F) are also well-studied in terms of the spin-valve effect \cite{Oh1997,Tagirov1999, Fominov2003,Moraru2006,Singh2007,Fominov2010,Zhu2010,Wu2012,Leksin2012,Banerjee2014,Jara2014,Singh2015,Kamashev2019,Westerholt2005,Deutscher1969,Gu2002,Gu2015}. Interestingly, theory for F1/F2/S systems by Fominov \textit{et al.} \cite{Fominov2010} shows the possibility of both "standard" and "reverse" spin-valve effect (when the critical temperature is lower in the P or AP configuration, respectively) due to constructive or destructive interference of the Cooper pair wave functions reflected from F1/F2 and F2/S interfaces. Besides, the $T_c$ dependence on the misorientation angle might be nonmonotonic with a minimum near $\phi=\pi/2$ \cite{Fominov2010,Wu2012,Leksin2012,Jara2014,Singh2015,Kamashev2019,Karminskaya2011} because of generation of the long-range triplet component which provides an additional way of superconductivity suppression in the case of non-collinear magnetizations. Research on the spin-valve-based devices is furthered \cite{Valls_book} because,
besides their great scientific interest, they have possible applications towards the creation of non-volatile magnetic memory
elements. The supercurrents passed through these devices can also be spin-polarized \cite{Wu2014,Halterman2015,Moen2017,Moen2018,Moen2020}, what can then lead to a low energy spin transfer torque that can be used to control the magnetization of nanoscale devices.

The dipolar stray fields and GHz frequency magnons
in F often cause parasitic detrimental influence in ferromagnet-based spintronic devices including superconducting spin valves.
Employing antiferromagnets (AFs) could significantly reduce these problems due to their zero net magnetization
and higher magnon frequencies \cite{Gomonay2014,Baltz2018,Jungwirth2016,Kamra2018,Brataas2020}. Simultaneously the zero net magnetization of AFs has long been considered to be an obstacle to the use of antiferromagnets in spintronic devices. However, it was recently shown that the superconducting spin-valve effect can also be realized in three-layer antiferromagnet/superconductor/antiferromagnet (AF/S/AF) structures despite the absence of macroscopic magnetization in the antiferromagnetic layers \cite{Johnsen_Kamra_2023}. In that work the AF/S/AF  spin valve with insulating antiferromagnets and fully compensated S/AF interfaces (that is, with zero interface magnetization) was theoretically investigated in Bogoliubov – de Gennes (BdG) framework. It may seem that such a system is invariant towards reversing the direction of the N\'eel vectors in one of the AFs and, consequently, there is no physical difference between parallel and antiparallel configurations. However, each of the S/AF interfaces generates triplet correlations called N\'eel triplets \cite{Bobkov2022}. The amplitude of these correlations changes its sign between the adjacent lattice sites in the superconductor in the same way as the magnetic order in the antiferromagnet does. The N\'eel triplets generated by the both S/AF interfaces may interfere constructively (destructively) inside the superconducting layer, thus suppressing superconductivity more (less) strongly. Therefore, the spin-valve effect is expected even in AF/S/AF heterostructures with fully compensated S/AF interfaces. 

In \cite{Johnsen_Kamra_2023} it was obtained that the critical temperature is sensitive to mutual orientation of the N\'eel vectors of the AFs and complete suppression of $T_c$, or the absolute spin-valve effect, is achievable. In the current work we supplement those results with a more detailed study of $T_c(\phi)$ behavior in various regimes. In particular, it is shown that similar to other important physical effects in S/AF hybrids, such as dependence of the critical temperature on impurity concentration \cite{Bobkov2022,Bobkov2023}, magnetic anisotropy of the critical temperature in the presence of spin-orbit coupling \cite{Bobkov2023_anisotropy} and dependence of the critical temperature on the canting angle \cite{Chourasia2023}, the physics of spin-valve effect is also very sensitive to the value of the chemical potential $\mu_S$ in the superconducting layer.  We observed an interesting and nontrivial dependence of $T_c$ on the misorientation angle between the N\'eel vectors of the AFs. In AF/S/AF structures there is some freedom in determination of the misorientation angle.  In \cite{Johnsen_Kamra_2023} the misorientation angle $\theta$ was defined as the angle between the N\'eel vectors of the closest to the S/AF interfaces antiferromagnetic layers. Here we focus on other aspects of  physics of AF/S/AF heterostructures and it is more convenient to choose a unified division of the entire AF/S/AF structure into two sublattices   and to define the misorientation angle $\phi$ as the angle between the
magnetizations of two antiferromagnets at the same sublattice, see Fig.~\ref{fig:setup} and the description of the model for further details of the definition.  In \cite{Johnsen_Kamra_2023} it was shown that near half-filling, that is at $\mu_S \approx 0$, the critical temperature is always lower for the parallel state corresponding to $\phi=0$ [$T_c(0) \equiv T_c^P$] than in the antipallel state corresponding to $\phi=\pi$ [$T_c(\pi) \equiv T_c^{AP}$]. Please notice that the parallel and antiparallel orientations are defined using the convention followed in the present manuscript.  It is explained by the fact that in this case the N\'eel triplets generated by the both interfaces are effectively summed up and strengthen each other inside the S layer. Here we demonstrate that if we move away from half-filling the opposite result $T_c^{P}>T_c^{AP}$ can be realized depending on the width of the S layer. We unveil the physical reason of this phenomenon, which is connected with the generation of finite-momentum N\'eel triplet pairs \cite{Bobkov2023_oscillatory}. Further, in \cite{Johnsen_Kamra_2023} it was indicated that the dependence $T_c(\phi)$ contains a contribution $\sim \sin^2 \phi$, which was ascribed to generation of equal-spin triplet correlations of conventional, not N\'eel, character. But near half-filling this contribution was found to be small. Here we investigate equal-spin contribution in more detail, provide its analytical description  and find parameter regions, where it is rather strong and results in the nonmonotonic dependence $T_c(\phi)$.  Finally, we discuss the dependence of the spin-valve effect on  the presence of impurities in the S layer and demonstrate that the difference $T_c^{P}-T_c^{AP}$ is suppressed by impurities due to the sensitivity of the N\'eel triplet correlations to impurities, but spin-valve effect produced by equal-spin pairs, which can be quantified by the difference $T_c^P+T_c^{AP}-2T_c(\phi=\pi/2)$, is not suppressed by impurities in the superconductor. 

The paper is organized as follows. In Sec.~\ref{quasiclassics} we present semi-analytical results obtained in the framework of the quasiclassical theory. Sec.~\ref{q_model} is devoted to description of the model of the AF/S/AF trilayer we study and the formalism of the two-sublattice quasiclassical theory applicable to S/AF hybrid structures. In Sec.~\ref{q_triplets} we present some analytical results and discuss the structure of triplet correlations responsible for the spin-valve effect in the AF/S/AF trilayer, while in Sec.~\ref{q_Tc}  the dependencies $T_c(\phi)$ obtained in the framework of our quasiclassical theory are provided and discussed. Sec.~\ref{BdG} is devoted to the presentation of numerical results obtained in the framework of the BdG approach. Sec.~\ref{bdg_method} briefly describes the method. In Sec.~\ref{bdg_chemical} we study the dependence of the spin-valve effect on the value of the chemical potential of the S layer, and Sec.~\ref{bdg_impurities} is devoted to the influence of impurities on the investigated effect. In Sec.~\ref{sec:materials} we describe the requirements for materials. Sec.~\ref{concl} contains the conclusions from our work. In the Appendix some technical details of the quasiclassical calculations are provided. 

\section{Quasiclassical theory: structure of triplet correlations and spin-valve effect}
\label{quasiclassics}

\subsection{Model and method}
\label{q_model}

\begin{figure}[tb]
	\begin{center}
		\includegraphics[width=100mm]{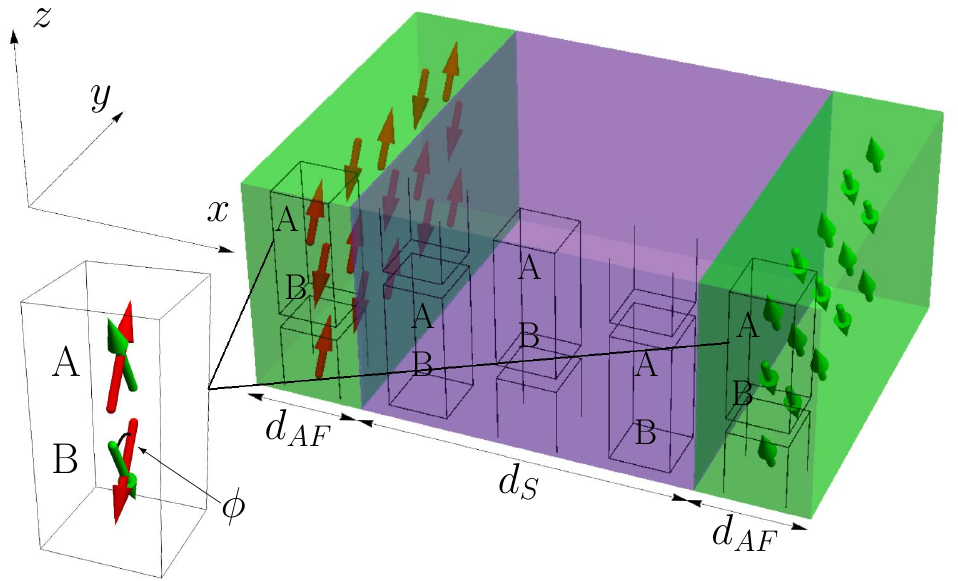}
		\caption{Sketch of the AF/S/AF system. Red and green arrows show N\'eel-type magnetizations of the AFs. The unified division into two sublattices with unit cells containing two sites belonging to $A$ and $B$ sublattices is also shown. The misorientation angle $\phi$ is defined as the angle between the magnetizations of two antiferromagnets at the same sublattice. It's important to note that this definition of the misorientation angle differs from the definition used in \cite{Johnsen_Kamra_2023}.}
        \label{fig:setup}
	\end{center}
\end{figure}

We consider an AF/S/AF trilayer system with fully compensated interfaces, depicted in Fig.~\ref{fig:setup}. A conventional $s$-wave singlet superconductor with thickness $d_S$ is sandwiched between two insulating antiferromagnets with the same thicknesses $d_{AF}$. For simplicity we study a two-dimensional system and employ periodic boundary conditions along the interfacial direction. We introduce a unified division into two sublattices for the entire AF/S/AF system. The misorientation angle $\phi$ is defined as the angle between the magnetizations of the both antiferromagnets at the same sublattice, see Fig.~\ref{fig:setup}. Please note that this definition of the misorientation angle differs from the definition used in \cite{Johnsen_Kamra_2023}. In that paper the misorientation angle $\theta$ was defined as the angle between the N\'eel vectors of the closest to the S/AF interfaces antiferromagnetic layers. The both definitions coincide for odd number of layers in the superconductor, but they differ by $\pi$ if the number of layers in the superconductor is even.

In order to obtain some analytical results on the structure of superconducting correlations in the S layer, including triplet ones, we  investigate the clean case $T_c(\phi)$ dependence in the framework of the two-sublattice quasiclassical theory \cite{Bobkov2022}. The unit cell containing two sites belonging to different sublattices $A$ and $B$ is shown in Fig.~\ref{fig:setup}. The Hamiltonian of the superconducting layer in the two-sublattice representation takes the form:
\begin{align}
\hat H= - &t \sum \limits_{\langle \bm{i}\bm{j}\nu \bar \nu\rangle ,\sigma} \hat \psi_{\bm{i} \sigma}^{\nu\dagger} \hat \psi_{\bm{j} \sigma}^{\bar \nu} + \sum \limits_{\bm{i},\nu } (\Delta_{\bm{i}}^\nu \hat \psi_{\bm{i}\uparrow}^{\nu\dagger} \hat \psi_{\bm{i}\downarrow}^{\nu\dagger} + H.c.) - \nonumber \\
&\mu_S \sum \limits_{\bm{i} \nu, \sigma} \hat n_{\bm{i}\sigma}^\nu  
+ \sum \limits_{\bm{i} \nu,\alpha \beta} \hat \psi_{\bm{i}\alpha}^{\nu \dagger} (\bm{h}_{\bm{i}}^\nu \bm{\sigma})_{\alpha \beta} \hat \psi_{\bm{i}\beta}^\nu.
\label{ham_2}
\end{align}
Here $\bm i$ is the radius vector of a unit cell as a whole, $\nu=A, B$ denotes the two sites in the unit cell corresponding to different sublattices, $\bar\nu=B(A)$ if $\nu=A(B)$. $\hat \psi_{\bm{i} \sigma}^{\nu \dagger} (\hat \psi_{\bm{i} \sigma}^\nu)$ is the creation (annihilation) operator for an electron with spin $\sigma$ at the site of the unit cell $\bm i = (i_x,i_y)^T$ and sublattice $\nu$. $x$- and $y$-axes are taken normal to the S/AF interfaces and parallel to them, respectively. $\langle \bm i \bm j \nu \bar \nu\rangle$ means summation over the nearest neighbors, $t$ denotes the hopping between adjacent sites, $\mu_S$ is the electron chemical potential, and $\hat n_{\bm i \sigma}^\nu = \hat \psi_{\bm i \sigma}^{\nu \dagger} \hat \psi_{\bm i \sigma}^\nu$ is the particle number operator at the site $(\bm i, \nu)$. $\Delta_{\bm i}^\nu$ and $\bm h_{\bm i}^\nu$ denote the on-site $s$-wave pairing and magnetic order parameter at the site $(\bm i, \nu)$, respectively. $\bm \sigma = (\sigma_x, \sigma_y, \sigma_z)^T$ is the vector of Pauli matrices in spin space. 

As we consider G-type antiferromagnets, the magnetic order parameter in the left  and right AFs can be taken in the form $\bm h_{\bm i}^{A(B)} = + (-) \bm h_{l,r}$. It is assumed that the antiferromagnetism is due to the localized electrons and the amplitude of the on-site magnetization is not influenced by the adjacent metal. Therefore, we do not calculate the magnetic order parameter self-consistently and consider it to have a constant value $h$ in the AF regions and zero value in the S region. For the on-site $s$-wave pairing in the S $\Delta_{\bm{i}}^A = \Delta_{\bm{i}}^B = \Delta_{\bm{i}}$.   $\Delta_{\bm i}$ is non-zero only in the superconductor.  

As all the parameters in the considered problem are slow functions of the lattice site spatial coordinate, we can introduce a continuous spatial variable $\bm R$ instead of the discrete index $\bm i$. Then the two-sublattice formalism allows us to describe the system with the quasiclassical Green's function  $\check g(\bm R, \bm p_F, \omega_m)$, which is an $8 \times 8$ matrix in the direct product of spin, particle-hole and sublattice spaces ($\bm p_F$ is the electron momentum at the Fermi surface, $\omega_m = \pi T(2m+1)$ is the fermionic Matsubara frequency). 

Let us choose the coordinate system $\bm e_x, \bm e_y, \bm e_z$, so that $\bm e_x$ is parallel to $\bm h_l\times \bm h_r$, $\bm e_z$ is parallel to $\bm h_l+\bm h_r$ and $x=0$ corresponds to the middle of the S layer. Then $\bm h_{l,r}$ can be written in the form
\begin{align}
    \bm h_l=h\left( \begin{array}{c}
0 \\
\sin (\phi/2) \\
\cos (\phi/2)
\end{array}\right),~~~\bm h_r=h\left( \begin{array}{c}
0 \\
-\sin (\phi/2) \\
\cos (\phi/2)
\end{array}\right). 
\end{align}
The quasiclassical Green's function in the superconductor obeys the following Eilenberger equation in the ballistic limit \cite{Bobkov2022}:
\begin{align}
    \Bigl[ \Bigl(i\omega_m \tau_z+\mu_S+\tau_z \check \Delta(\bm R)-\bm h(x)\bm\sigma\rho_z \tau_z a\Bigr)\rho_x, \check g\Bigr]+i \bm v_F \bm \nabla \check g=0, 
\label{Eilen}
\end{align}
where  $\bm h(x)=\bm h_{l}\delta(x+d_S/2)+\bm h_{r}\delta(x- d_S/2)$, $\bm v_F$ is the Fermi velocity for the trajectory $\bm p_F$, $\tau_i$ and $\rho_i$ are Pauli matrices in particle-hole and sublattice spaces, respectively, $\check \Delta(\bm R) = \Delta(\bm R) \tau_x$. The term proportionate to $\bm h_{l,r}\bm\sigma a\delta(x\pm d_S/2)$  accounts for the exchange field at the left and right S/AF interfaces $x=\mp d_S/2$, $a$ is the lattice constant of the superconductor along the $x$-direction.

As we consider the system translationally invariant along the S/AF interfaces, the Green's function depends only on the $x$ coordinate, normal to the interfaces. Then the term $i \bm v_F \bm \nabla \check g$ in (\ref{Eilen}) is reduced to $i v_{F,x} d\check g/dx$. We define the Green's functions corresponding to the trajectories, incident to the right S/AF interface and reflected from it, respectively, as $\check g_+ (x)\equiv  \check g (x, v)$ and $ \check g_- (x)\equiv \check g (x, -v)$, where for brevity we introduce the notation $v \equiv |v_{F,x}|$. In addition to the Eilenberger equation (\ref{Eilen}), the quasiclassical Green's function for a given trajectory $\check g_{+(-)} (x)$ obeys the normalization condition
\begin{align}
     \left[\check g_{+(-)}(x)\right]^2=1
\label{norm}
\end{align}
and the boundary conditions at the S/AF interfaces $x=\mp d_S/2$, which, due to the symmetrically chosen coordinate system, are reduced to one boundary condition at one of the interfaces (for the following calculations we take the right interface $x= d_S/2$).

In the problem under consideration the S layer width $d_S$ is assumed to be much smaller than the superconducting coherence length $\xi_S = v_F/2\pi T_{c0}$, where $T_{c0}$ is the superconducting bulk critical temperature. Therefore we can consider the superconducting order parameter in the S region spatially constant: $\Delta(\bm R)\approx \Delta$. 
The explicit structure of the Green's function in the particle-hole space takes the form
\begin{align}
\check g = 
\left(
\begin{array}{cc}
\hat g & \hat f \\
\hat {\tilde f} & \hat {\tilde g}
\end{array}
\right)_{\tau},
\label{gf_linearized}    
\end{align}
where all the components are $4\times4$ matrices in the direct product of spin and sublattice spaces. As we study the system at temperatures close to the critical temperature, the Eilenberger equation (\ref{Eilen}) can be linearized with respect to $\Delta/T_c$.  The diagonal components $\hat g$ and $\hat {\tilde g}$ are to be calculated in the normal state of the superconductor and the anomalous components $\hat f$ and $\hat {\tilde f}$ are of the first order with respect to $\Delta/T_c$.

For the normal state Green's function 
\begin{align}
\check g^N = 
\left(
\begin{array}{cc}
\hat g & 0 \\
0 & \hat {\tilde g}
\end{array}
\right)_{\tau}
\label{g_normal}    
\end{align}
the Eilenberger equation (\ref{Eilen}) takes the form
\begin{align}
    \left[ (i\omega_m \tau_z+\mu_S-\bm h(x)\bm\sigma\rho_z \tau_z a)\rho_x, \check g^{N}\right]+i v_{F,x}  \dfrac{d}{dx}\check g^{N}=0 
\label{Eilen_g}
\end{align}
and leads to the following boundary condition at $x=d_S/2$ (see the Appendix):
\begin{align}
    \delta \check g^{N}\equiv \check g^{N}_-(x=d_S/2)-\check g^{N}_+(x=d_S/2)=\frac{2\tau_z}{v}[a\bm h_r \bm \sigma\rho_y, \frac{\hat g_+(x=d_S/2)+\hat g_-(x=d_S/2)}{2}]
    \label{boundary_g}.
\end{align}
The detailed calculation of $\check g^{N},$ as well as some extra conditions on it resulting from symmetry, are presented in the Appendix.
Let us define the following expansions of $\hat g, \hat f, \hat {\tilde f}$ and $\hat {\tilde g}$ over the Pauli matrices in the sublattice space and over the direct product of Pauli matrices in spin and sublattice spaces: 
\begin{align}
\hat g=\sum\limits_\alpha\rho_\alpha \hat g_\alpha,~~~\hat g=\sum\limits_{\alpha,\beta}\sigma_\alpha\rho_\beta g_{\alpha\beta},
\label{exp}
\end{align}
and similarly for $\hat f, \hat {\tilde f}$ and $\hat {\tilde g}$. The symmetry also gives us the following relation between $\check g_-(x)$ and $\check g_+(-x)$, see Appendix for details of the derivation:
\begin{align}
\begin{cases}
    \check g_{0\alpha,-}(x)=\check g_{0\alpha,+}(-x)\\
    \check g_{x\alpha,-}(x)=-\check g_{x\alpha,+}(-x)\\
    \check g_{y\alpha,-}(x)=-\check g_{y\alpha,+}(-x)\\
    \check g_{z\alpha,-}(x)=\check g_{z\alpha,+}(-x),
    \end{cases}
    \label{symm}
\end{align}
where 
\begin{align}
\check g_{\beta\alpha, +(-)} \equiv
\left(
\begin{array}{cc}
g_{\beta\alpha, +(-)} & f_{\beta\alpha, +(-)} \\
{\tilde f}_{\beta\alpha, +(-)} & {\tilde g}_{\beta\alpha, +(-)}
\end{array}
\right).   
\end{align}
Therefore in the following text we write the expressions only for $\check g_+(x)\equiv \check g(x)$. The solution for the particle component of the normal state Green's function takes the form
\begin{align}
     \hat g={\rm sgn} ~\omega_m\left[g_{0x}\sigma_0\rho_x  +g_{x0}\sigma_x\rho_0  +B\sigma_y(i\rho_y \sinh \kappa x + \rho_z ~\textrm{sgn}~ v_{F,x} \cosh \kappa x)+\right. \nonumber\\ \left.A\sigma_z(i \rho_z ~\textrm{sgn}~ v_{F,x} \sinh \kappa x-\rho_y\cosh \kappa x)\right],
\label{g_N1}
\end{align}
where $\kappa=2(i\mu_S-\omega_m)/v$ and the coefficients $g_{0x}, g_{x0}, A$ and $B$ are found from the normalization condition (\ref{norm}) and the boundary condition (\ref{boundary_g}) and take the form, see Appendix for the derivation:
\begin{align}
    g_{0x}=\dfrac{1}{\sqrt{(1+\gamma_a^2)(1+\gamma_b^2)}},\nonumber \\    g_{x0}=\dfrac{\gamma_a\gamma_b}{\sqrt{(1+\gamma_a^2)(1+\gamma_b^2)}},\nonumber \\
    A=\dfrac{\gamma_a}{\sqrt{(1+\gamma_a^2)(1+\gamma_b^2)}},\nonumber \\
    B=\dfrac{\gamma_b}{\sqrt{(1+\gamma_a^2)(1+\gamma_b^2)}},
\end{align}
where $\gamma_a=2ah_z/[v\sinh (\kappa d_S/2)], \gamma_b=2iah_y/[v\cosh (\kappa d_S/2)]$, $h_{y,z}\equiv (\bm h_l)_{y,z}$.  
The hole component $\hat {\tilde g}$ is obtained from (\ref{g_N1}) by the relation
\begin{align}
     \hat {\tilde g}(\omega_m, \bm h_{r}, \mu_S)=\hat g(-\omega_m, -\bm h_{r}, \mu_S).
\label{g_N2}
\end{align}

The anomalous Green's function $\hat f$ is found from the linearized with respect to $\Delta/T_c$ Eilenberger equation
\begin{align}
     \{\left(i\omega_m-\bm h(x)\bm\sigma\rho_z a\right)\rho_x, \hat f\}+\mu_S [\rho_x, \hat f]+\Delta(\rho_x\hat {\tilde g}-\hat g\rho_x)+iv_{F,x}\dfrac{d}{dx}\hat f=0
\label{Eilen_f}
\end{align}
where $\{F_1,F_2\}=F_1F_2+F_2F_1$ means anticommutator. And the boundary condition
\begin{align}
     \delta \hat f\equiv \hat f_-(x=d_S/2)-\hat f_+(x=d_S/2)=\dfrac{2}{v}\{a\bm h_r \bm \sigma\rho_y, \frac{\hat f_+(x=d_S/2)+\hat f_-(x=d_S/2)}{2}\}.
\label{boundary_f}
\end{align}
The details of the calculation of $\hat f$ are presented in the Appendix. It is convenient to define the following $2\times 2$ matrices in the sublattice space: 
\begin{align}
     \hat G_0=\Delta(\hat g_0-\hat {\tilde g}_0), \nonumber\\
     \hat G_x=\Delta(\hat g_x-\hat {\tilde g}_x),\nonumber\\
     \hat G_y=-i\Delta(\hat g_y+\hat {\tilde g}_y),\nonumber\\
     \hat G_z=i\Delta(\hat g_z+\hat {\tilde g}_z).
\label{G}
\end{align}
Let us write the following expansions up to the first order with respect to $x$ for the matrices (\ref{G}): $\hat G_\beta=\sum\limits_\alpha \sigma_\alpha G_{\alpha\beta}, ~\hat G_{y,z}=\hat G_{y,z}^0+\hat G'_{y,z}x$. Then in the linear order with respect to $x/\xi_S$ the solution for the anomalous Green's function takes the form $\hat f=\sum\limits_{\alpha,\beta}\sigma_\alpha\rho_\beta f_{\alpha\beta}$ with the following non-zero components:
\begin{align}
     f_{00}=A_0\frac{2\omega_m}{v}x~\textrm{sgn}~v_{F,x}, \nonumber\\
     f_{0x}=-A_0+\frac{G_{0x}}{2i\omega_m},\nonumber\\
     f_{xx}=-B_x\frac{2\omega_m }{v}x,\nonumber\\
     f_{yy}=\Bigl(C_y\frac{2\mu_S}{v}+\frac{G'_{yy}}{2i\mu_S}\Bigr)x,\nonumber\\
     f_{yz}=\textrm{sgn}~v_{F,x}\Bigl(C_y+\frac{1}{2i\mu_S}\Bigl(-G_{yz}^0+\frac{v}{2\mu_S}G'_{yy}\Bigr)\Bigr),\nonumber\\ f_{zy}=D_z+\frac{1}{2i\mu_S}\Bigl(G_{zy}^0+\frac{v}{2\mu_S}G'_{zz}\Bigr),\nonumber\\ f_{zz}=-x~\textrm{sgn}~v_{F,x}\Bigl(D_z\frac{2i\mu_S}{v}-\frac{G'_{zz}}{2i\mu_S}\Bigr),     
\label{f}
\end{align}
where the coefficients $A_0, B_x, C_y$ and $D_z$ are obtained from the boundary condition (\ref{boundary_f}) and explicitly written in the Appendix.

The critical temperature for each value of the misorientation angle $\phi$ is calculated from the self-consistency equation 
\begin{align}
\Delta(x) = \int \frac{d\Omega}{4\pi} i \pi \lambda  T_c \sum \limits_{\omega_m} f_{0x}(x) ,
\label{Tc}    
\end{align}
where $\int \frac{d\Omega}{4\pi}$ means averaging over the Fermi surface and $\lambda$ is the dimensionless coupling constant. The amplitude of on-site singlet correlations $f_{0x}$ is spatially constant in the first order with respect to $x$.

\subsection{Structure of triplet correlations}
\label{q_triplets}

Now our goal is to discuss the results for the dependence $T_c(\phi)$ in the framework of the quasiclassical theory. First, let us analyze the expression for the full anomalous Green's function in order to find out the types of triplet correlations that different parts of this expression correspond to. It can be written in the form:
\begin{align}
    \hat f=\hat f_0 \sigma_0+\hat f_l \bm{h}_{\mathrm{eff},l} \bm \sigma+\hat f_r \bm{h}_{\mathrm{eff},r} \bm \sigma+\hat f_{\mathrm{cross}} ({\bm h}_{\mathrm{eff},l} \times \bm{h}_{\mathrm{eff},r}) \bm \sigma,
\end{align}
where the amplitudes $\hat f_i$ are matrices in the sublattice space and we introduce the combination ${\bm h}_{\mathrm{eff},l(r)} = \bm h_{l(r)} a/d_S$, which is called effective exchange field and physically corresponds to the averaging of the interface exchange field over the whole width $d_S$ of the S layer. Further we also use the absolute value of the effective exchange field $h_{\mathrm{eff}} \equiv ha/d_S$. The amplitudes $\hat f_l$  and $\hat f_r$ contain only $\rho_{y,z}$ contributions, which means they are correlations of the N\'eel type. On the contrary, the cross product amplitude $\hat f_{\mathrm{cross}}$ contains only $\rho_{x}$ contribution and, consequently, corresponds to conventional on-site equal-spin even-momentum odd-frequency triplet correlations. The cross product correlations are maximal at $\phi \approx \pi/2$. To the leading order with respect to $d_S/\xi_S$ the amplitude $\hat f_{\mathrm{cross}}$ takes the form:
\begin{align}
    \hat f_{\mathrm{cross}}=-\frac{2\Delta x d_S \hat\rho_x}{v^2}\left(\frac{{\rm sgn}\omega_m}{\sqrt{(\omega_m-i\mu_S)^2+4h_{\mathrm{eff,z}}^2}}-\frac{{\rm sgn}\omega_m}{\sqrt{(\omega_m+i\mu_S)^2+4h_{\mathrm{eff,z}}^2}}\right),
    \label{f_cross}
\end{align}
where $h_{\mathrm{eff},z} = h_{\mathrm{eff}} \cos \phi/2$. It is seen that $\hat f_{\mathrm{cross}}$ is of the first order with respect to $d_S/\xi_S$ and, therefore, is not pronounced in structures with thin S layers. This behavior differs from the behavior of $\hat f_{l,r}$, which remain finite at $d_S/\xi_S \to 0$ with $\hat f_l = \hat f_r$, what means that for the thin S layers the N\'eel correlations are only determined by the vector sum of the effective exchange fields produced by the both S/AF interfaces, that is indeed the resulting effective exchange field $\propto \cos \phi/2$, as it was reported for ferromagnets \cite{DEGENNES1966}. 

Moreover, in the framework of the quasiclassical approximation $\hat f_{\mathrm{cross}}$ is an odd function of $\mu_S$ and consequently $\hat f_{\mathrm{cross}} = 0$ at $\mu_S = 0$. It is in agreement with the smallness of the $\sin^2 \phi$ contribution in \cite{Johnsen_Kamra_2023}, where the case $\mu_S=0$ was considered and, therefore, the equal-spin triplet correlations could contribute to this term only beyond the quasiclassical approximation.

\subsection{Dependence of the critical temperature on the misorientation angle}
\label{q_Tc}

\begin{figure}[tb]
	\begin{center}
		\includegraphics[width=85mm]{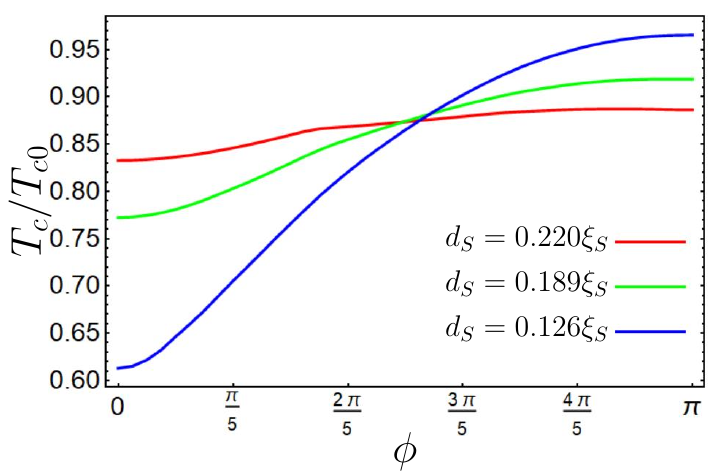}
		\caption{$T_c(\phi)$ for the AF/S/AF structure in the framework of the quasiclassical approach for fixed $h_{\mathrm{eff}}=T_{c0}$ and different $d_S$. $\mu_S =T_{c0} $. $T_c$ is measured in units of the superconducting bulk critical temperature $T_{c0}$ throughout the text.}
         \label{fig:Quasfixedh}
	\end{center}
\end{figure}

\begin{figure}[tb]
	\begin{center}
		\includegraphics[width=85mm]{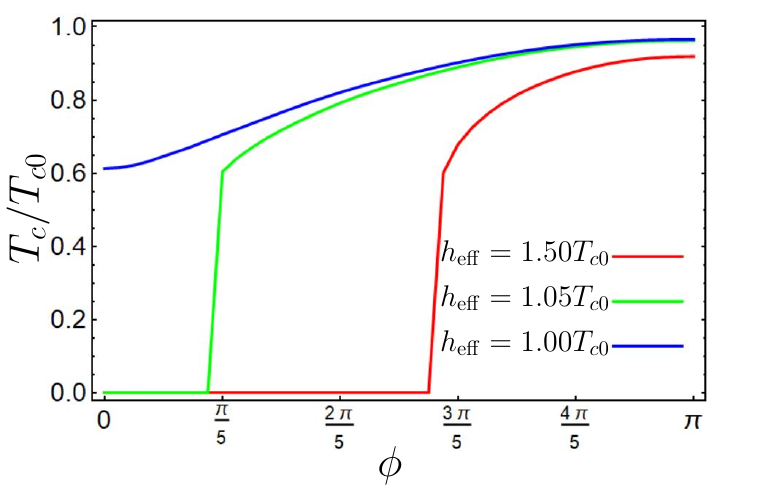}
		\caption{$T_c(\phi)$ for a fixed $d_S=0.126\xi_S$ and different $h_{\mathrm{eff}}$. For larger values of $h_{\mathrm{eff}}$ the spin-valve effect is absolute. $\mu_S =T_{c0}$.}
    \label{fig:Quasabsolute}	
 \end{center}
\end{figure}

Figs.~\ref{fig:Quasfixedh} and \ref{fig:Quasabsolute} demonstrate the dependence of the critical temperature on the misorientation angle $\phi$. In the limit of thin S layer the influence of the N\'eel exchange field of the AFs on the S layer is determined by the effective exchange field $h_{\mathrm{eff}} = ha/d_S$. That is, the narrower the S layer is, the stronger effective exchange it feels due to the proximity effect with the  antiferromagnets. Fig.~\ref{fig:Quasfixedh} shows $T_c(\phi)$ for a fixed  $h_{\mathrm{eff}}$. The spin-valve effect is well-pronounced in all the results in this subsection. The suppression of the critical temperature is maximal at $\phi=0$ according to the dependence of the effective exchange field on the misoriention angle, which is $2 h_{\mathrm{eff}}\cos(\phi/2)$. Thus, here we reached agreement with the expected relation $T_c(\phi=0)<T_c(\phi=\pi)$.

Fig.~\ref{fig:Quasfixedh} shows that the valve effect is reduced for larger superconducting width $d_S$: the $T_c$ suppression becomes weaker at $\phi = 0$ and increases at $\phi=\pi$. This general trend is physically clear. From physical considerations it follows that in the limit $d_S \gg \xi_S$ the valve effect should disappear because the two S/AF interfaces do not feel each other and the superconductivity suppression at each of them does not depend on the direction of the N\'eel vector. On the contrary, for the thinnest  superconductors $d_S \ll \xi_S$ the critical temperature is not suppressed from its bulk value at $\phi=\pi$ due to the exact compensation of the N\'eel triplets generated at the both interfaces. Therefore, at $\phi = \pi$ the maximal value of the critical temperature is reached for the thinnest S layers with $d_S \ll \xi_S$, at larger $d_S$ the critical temperature is lower due to the uncorrelated superconductivity suppression by the N\'eel exchange field at the S/AF interfaces separately.

Another feature of the results presented in Fig.~\ref{fig:Quasfixedh} is the following. For short superconducting interlayers (blue and green curves) the $T_c(\phi)$ dependence is smooth, but for wider $d_S$ the distortion in the vicinity $\phi=\pi/2$ is seen. This is the manifestation of the  cross product term $\sim \bm h_l \times \bm h_r$, which influences singlet correlations with the maximal effect  at $\phi = \pi/2$.  At $d_S/\xi_S \to 0$ this term disappears, and all the effect of the antiferromagnets on the S layer is described by the total effective exchange field $2 h_{\mathrm{eff}} \cos \phi/2$. For this reason in the framework of our analytical first-order approximation with respect to $d_S/\xi_S$ it is not possible to obtain more pronounced signatures of these correlations (in the form of dips): higher orders of $d_S/\xi_S$ need to be taken into account. We demonstrate more pronounced dip features at the dependence $T_c(\phi)$ in the framework of the BdG approach below. From our quasiclassical theory it follows that at $h_{\mathrm{eff}} \ll T_c$ the dependence $T_c(\phi)$ takes the form $T_c=T_{c,\parallel}+ \Delta T_{c,\parallel} \cos \phi + \Delta T_{c,\perp} \sin^2 \phi$, as it was phenomenologically proposed in \cite{Johnsen_Kamra_2023}. Here $T_{c,\parallel} = (T_c(0)+ T_c(\pi))/2$, $\Delta T_{c,\parallel} = (T_c(0)- T_c(\pi))/2$, $\Delta T_{c,\perp}=T_c(\pi/2)-T_{c,\parallel}$. However, at higher $h_{\mathrm{eff}}$ plots in Fig.~\ref{fig:Quasfixedh} show a clear deviation from this formula, due to the fact that the contributions of higher powers of $(\bm h_{\mathrm{eff},l} \bm h_{\mathrm{eff},r})$ and $h_{\mathrm{eff},z(y)}$ become more significant.

Fig.~\ref{fig:Quasabsolute} demonstrates that for the system under consideration the absolute spin-valve effect, that is the full suppression of superconducting state for a range of misorientation angles, is also possible. This result is in agreement with the one presented in \cite{Johnsen_Kamra_2023}.

\section{Bogoliubov – de Gennes approach: dependence of the spin-valve effect on chemical potential and impurities}
\label{BdG}

\subsection{Method}
\label{bdg_method}

The system is described by a tight-binding Hamiltonian Eq.~(\ref{ham_2}), but for the Bogoliubov – de Gennes calculations \cite{Zhu2016} it is convenient to get rid of sublattices:
\begin{align}
\hat H= - t \sum \limits_{\langle \bm{i}\bm{j} \rangle, \sigma} \hat \psi_{\bm{i} \sigma}^\dagger \hat \psi_{\bm{j} \sigma} + \sum \limits_{ \bm{i}} (\Delta_{\bm{i}} \hat \psi_{\bm{i}\uparrow}^\dagger \hat \psi_{\bm{i}\downarrow}^\dagger + H.c.) - \mu \sum \limits_{\bm{i}, \sigma} \hat n_{\bm{i}\sigma} + \sum \limits_{\bm{i},\alpha \beta} \hat \psi_{\bm{i}\alpha}^\dagger (\bm{h}_{\bm{i}} \bm{\sigma})_{\alpha \beta} \hat \psi_{\bm{i}\beta} .
\label{ham}
\end{align}
Here $\hat \psi_{\bm{i} \sigma}^\dagger (\hat \psi_{\bm{i} \sigma})$ is the creation (annihilation) operator for an electron with spin $\sigma$ at the site with the radius vector $\bm i = (i_x,i_y)^T$. $\langle \bm i \bm j \rangle$ means summation over the nearest neighbors, $\hat n_{\bm i \sigma} = \hat \psi_{\bm i \sigma}^\dagger \hat \psi_{\bm i \sigma}$ is the particle number operator at the site $\bm i$. $\Delta_{\bm i}$ and $\bm h_{\bm i}$ denote the on-site $s$-wave pairing and magnetic order parameter at the site $\bm i$, respectively. The N\'eel exchange field can be taken in the form $\bm h_{\bm i,l} =  (-1)^{i_x+i_y} \bm h_l$ and $\bm h_{\bm i,r} =  (-1)^{i_x+i_y} \bm h_r$ in the left and the right AF regions, respectively. It is worth noting that in the present section the considered AF insulator is also described by hopping hamiltonian Eq.~(\ref{ham}). As a result it has a finite bandgap such that there is a leakage of the electronic wavefunctions into the AFs. In fact, the wavefunctions penetrate to 2-3 sites. This is in contrast with the ideal antiferromagnetic insulators considered in Ref. \cite{Johnsen_Kamra_2023} and the quasiclassical theory above.

We diagonalize the Hamiltonian (\ref{ham}) by the Bogoliubov transformation:
\begin{align}
\hat \psi_{\bm i\sigma}=\sum\limits_n \left(u^{\bm i}_{n\sigma}\hat b_n+v^{\bm i*}_{n\sigma}\hat b_n^\dagger\right), 
\label{bogolubov}
\end{align}
where $\hat b_n^\dagger (\hat b_n)$ are the creation (annihilation) operators of Bogoliubov quasiparticles. Then the resulting Bogoliubov – de Gennes equations take the form:
	\begin{align}
	-\mu u^{\bm i}_{n,\sigma}-t\sum \limits_{\bm j \in \langle \bm i \rangle}u_{n,\sigma}^{\bm j} 
	+ \sigma \Delta_{\bm i} v^{\bm i}_{n,-\sigma}+(\bm {h}_{\bm i} \bm{\sigma})_{\sigma\alpha}u_{n,\alpha}^{\bm i} & = \varepsilon_n u_{n,\sigma}^{\bm i} \nonumber \\  
	-\mu v^{\bm i}_{n,\sigma}-t \sum \limits_{\bm j \in \langle \bm i \rangle}v_{n,\sigma}^{\bm j}
	+ \sigma \Delta_{\bm i}^* u^{\bm i}_{n,-\sigma}+(\bm{h}_{\bm i}\bm{\sigma}^*)_{\sigma\alpha}v_{n,\alpha}^{\bm i} & = -\varepsilon_n v_{n,\sigma}^{\bm i}, 
	\label{eq:bdg}
	\end{align}
where $\bm j \in \langle \bm i \rangle$ means summation over the nearest neighbors $\bm j$ of the site $\bm i$ and $\varepsilon_n$ are the eigen-state energies of the Bogoliubov quasiparticles. The superconducting order parameter in the S layer is calculated self-consistently:
\begin{align}
\Delta_{\boldsymbol{i}}= g\langle\hat \psi_{\bm{i} \downarrow} \hat \psi_{\bm{i} \uparrow} \rangle =  g \sum\limits_n \left(u_{n,\downarrow}^{\bm i} v_{n,\uparrow}^{\bm i*}(1-f_n)+u_{n,\uparrow}^{\bm i} v_{n,\downarrow}^{\bm i*}f_n\right),
\end{align}
where $g$ is the coupling constant. The quasiparticle distribution function is assumed to be the equilibrium Fermi distribution $f_n = \langle b_n^\dagger b_n \rangle = 1/(1+e^{\varepsilon_n/T})$. 

Since the simulations can only deal with a small number of lattice sites, it is typical in the BdG approach to employ rescaled parameters that respect the hierarchy of energy scales~\cite{Zhu2016}, but do not completely mimic an actual material. For example, in a realistic material, we expect $t \sim 1000T_{c0}$. However, this choice would necessitate the numerical evaluation of eigenenergies with an impossibly high precision. Thus, we choose $T_{c0} \sim (0.01 - 0.1) t$ and $h_{\mathrm {eff}} = ha/d_S \sim T_{c0}$. Our analysis then faithfully reproduces realistic systems in which the essential physics depends on the relative strengths of $h$ and $T_{c0}$, but not on $t$ in consistence with the quasiclassical limit $t \gg T_{c0}$.

\subsection{Dependence of the spin-valve effect on the chemical potential}
\label{bdg_chemical}

In this subsection, applying the described technique, we investigate the influence of  the chemical potential on  the dependence $T_c(\phi)$. In \cite{Johnsen_Kamra_2023} it was demonstrated that at small $h \ll (\mu_S,t)$ (what means that $h_{\mathrm{eff}}/\mu_S < h/\mu_S$ is also small) the spin-valve effect is negligible with respect to the effect observed at half-filling $\mu_S = 0$. This is due to the fact that away from half-filling, that is at $\mu_S \gtrsim T_{c0}$, the amplitude of the N\'eel triplets, mediating the spin-valve effect, is $\sim h_{\mathrm{eff}}/\mu_S$ \cite{Bobkov2023}. However, in relation to real systems the both cases $h>\mu_S$ and $h<\mu_S$ can be realized. Therefore the parameter region $h_{\mathrm{eff}} \sim \mu_S$ looks reasonable and  also requires investigation. Here we demonstrate that the spin-valve effect persists in this regime. Moreover, in contrast to the case $\mu_S=0$, where the relation $T_c(0)<T_c(\pi)$ always holds, at larger $\mu_S$ the relation between $T_c(0)$ and $T_c(\pi)$ can be opposite. Below we demonstrate this result and discuss its physical reasons. 

\begin{figure}[tb]
	\begin{center}
		\includegraphics[width=85mm]{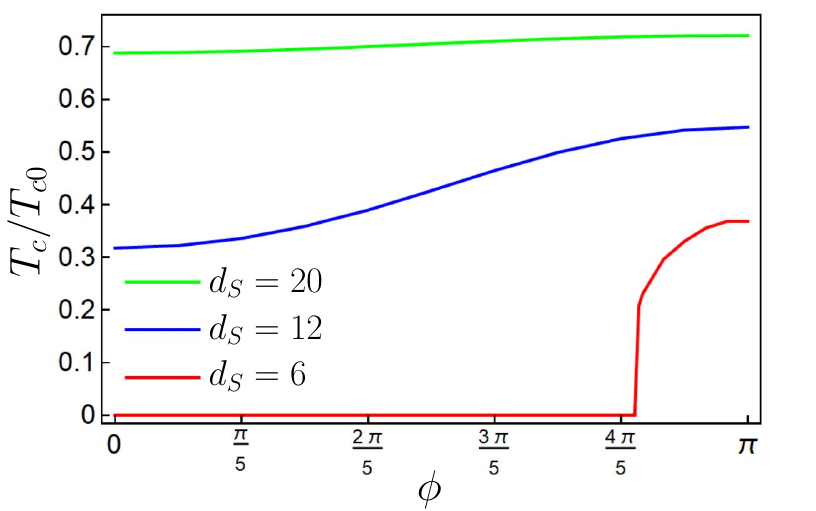}
		\caption{$T_c(\phi)$ for the AF/S/AF structure in the framework of BdG approach at half-filling $\mu_S=0$. Different curves correspond to different $d_S$, $d_{AF}=4$ (all widths are measured in the number of monolayers), $\mu_{AF}=0$, $h=0.5t$, $T_{c0}=0.07t$. }
        \label{fig:Tcphi_0}
	\end{center}
\end{figure}

\begin{figure}[tb]
	\begin{center}
		\includegraphics[width=85mm]{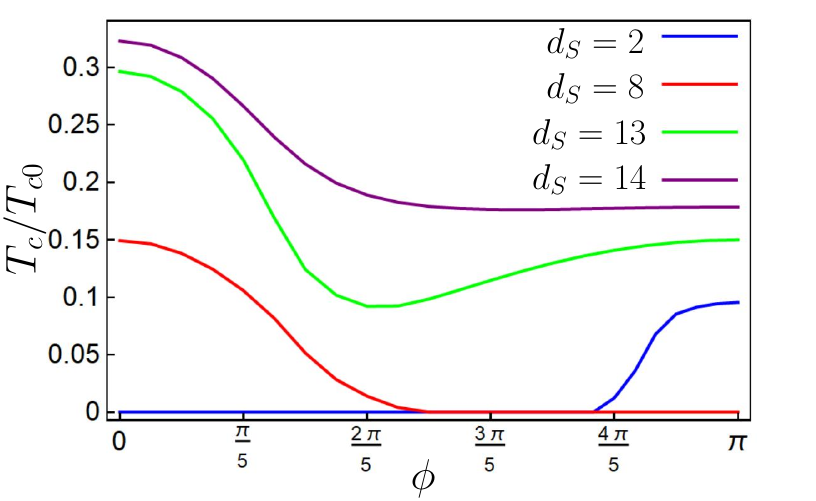}
		\caption{$T_c(\phi)$ for the AF/S/AF structure in the framework of BdG approach at $\mu_S=0.2 t$. Different curves correspond to different $d_S$, $d_{AF}=4$, $\mu_{AF}=0$, $h=0.5t$, $T_{c0}=0.07t$. }
        \label{fig:Tcphi_l}
	\end{center}
\end{figure}

\begin{figure}[tb]
	\begin{center}
		\includegraphics[width=85mm]{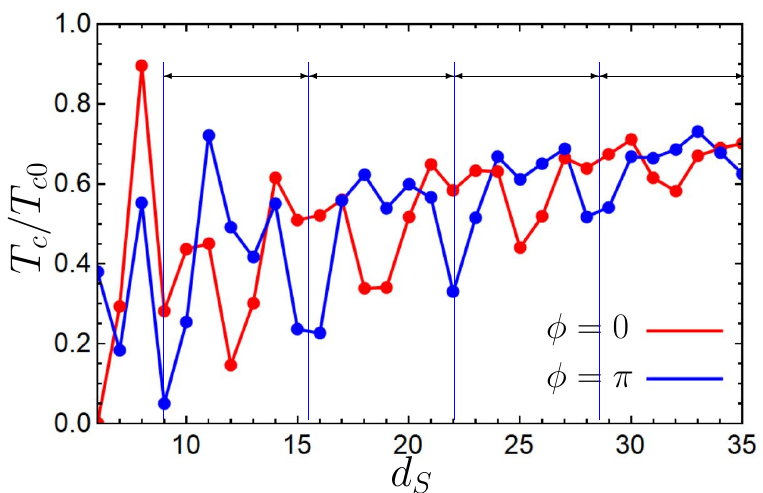}
		\caption{$T_c(0)$ and $T_c(\pi)$ as functions of $d_S$ at $\mu_S=0.9 t$.  $d_{AF}=4$, $\mu_{AF}=0$, $h=t$, $T_{c0}=0.03t$. $h_{\mathrm{eff}} = ha/d_S \ll \mu_S$, consequently for this set of parameters $L_{\mathrm{osc}} =  \pi v_F/\mu_S \approx 7$, what is in agreement with the data: four periods (minima $T_c(d_S)$ for $\phi=\pi$) are shown on the plot by vertical blue lines, so $L_{\mathrm{osc}}=\frac{35-9}{4}=6.5$. The small local minima inside some periods are caused by Friedel oscillations due to unrealistic small number of sites used in the BdG aproach.}
        \label{fig:Tc_osc}
	\end{center}
\end{figure}

In this subsection we assume no impurities in the S layer. Then the translational invariance allows us to consider the cluster infinite along the interfacial direction and solve a 1D problem for a system with the width $W=d_{AF}+d_S+d_{AF}$ along the $x$-direction. The dependencies $T_c(\phi)$ at $\mu_S = 0$ and $\mu_S = 0.2 t$ are shown in Figs.~\ref{fig:Tcphi_0} and \ref{fig:Tcphi_l}, respectively. For the data presented in these figures $\xi_S = v_F/2\pi T_c \approx 6$ monolayers. We can observe that at $\mu_S = 0$ the results are in agreement with our quasiclassical results presented in the previous section and also they are in agreement with \cite{Johnsen_Kamra_2023}. The relation $T_c(0)<T_c(\pi)$ is fulfilled for all considered values of $d_S$. However, as it is seen from Fig.~\ref{fig:Tcphi_l}, at larger $\mu_S$ the relation between $T_c(0)$ and $T_c(\pi)$ depends on the value of $d_S$ and opposite cases can be realized. The reason is the finite momentum, acquired by the N\'eel triplet Cooper pairs \cite{Bobkov2023_oscillatory} in systems with broken translational invariance via the Umklapp scattering processes at the S/AF interfaces. Due to the finite momentum of the N\'eel triplet Cooper pairs their wave function oscillates in the S layer with the period $L_{\mathrm{osc}} = \pi v_F/|\mu_S|$.  Depending on the width of the S layer $d_S$ the N\'eel triplets generated by the opposite S/AF interfaces can interfere constructively or destructively in the S layer, which manifests itself in the oscillating behavior of the resulting N\'eel triplet amplitude for a given $\phi$ upon varying $d_S$. This physical picture is further supported by the demonstration of the dependence of $T_c(0,\pi)$ on $d_S$ presented in Fig.~\ref{fig:Tc_osc}. The oscillations of the difference $T_c(\pi)-T_c(0)$ with the period $L_{\mathrm{osc}}$ are clearly seen.

The other feature worth mentioning is the nonmonotonicity of the curves in Fig.~\ref{fig:Tcphi_l}. The dip in the critical temperature at $\phi$ close to $\pi/2$ can be explained by generating of equal-spin triplet correlations determined by $\bm h_l \times \bm h_r$. These correlations are not of sign-changing N\'eel type and are usual equal-spin triplet correlations. The dip can be clearly seen for $d_S=13$ monolayers. For lower values of the S width the equal-spin triplet correlations are too weak to result in the pronounced dip feature because they vanish at $d_S/\xi_S \ll 1$, see Eq.~(\ref{f_cross}). For some higher values of $d_S$ the influence of the equal-spin triplet correlations can be  superimposed by the interference effects due to the finite-momentum N\'eel triplet pairing, which also mask their effect. 

\subsection{Dependence of the spin-valve effect on impurities}
\label{bdg_impurities}

Now we discuss the influence of impurities in the S region on the $T_c(\phi)$ dependence and spin-valve effect. The impurities are modelled as random changes of the chemical potential $\mu_S$ at each site of the superconductor:
\begin{align}
    \mu_i=\mu_S+\delta \mu_i, ~~~~~\delta \mu_i \in [-\delta\mu,\delta\mu],
\end{align}  
which break the translational invariance along the interface. For this reason we now investigate the 2D cluster with the previously defined width $W$ and finite length $L$ under periodic boundary conditions along the $y$-direction. It has to be noted that realistic samples should contain a much larger number of sites than it is possible to use in our calculations without making them incredibly time-consuming. In order to reasonably simulate this in the framework of our approach, we average the results for the critical temperature over 5-10 realizations of the impurity pattern.  


\begin{figure}[tb]
	\begin{center}
		\includegraphics[width=85mm]{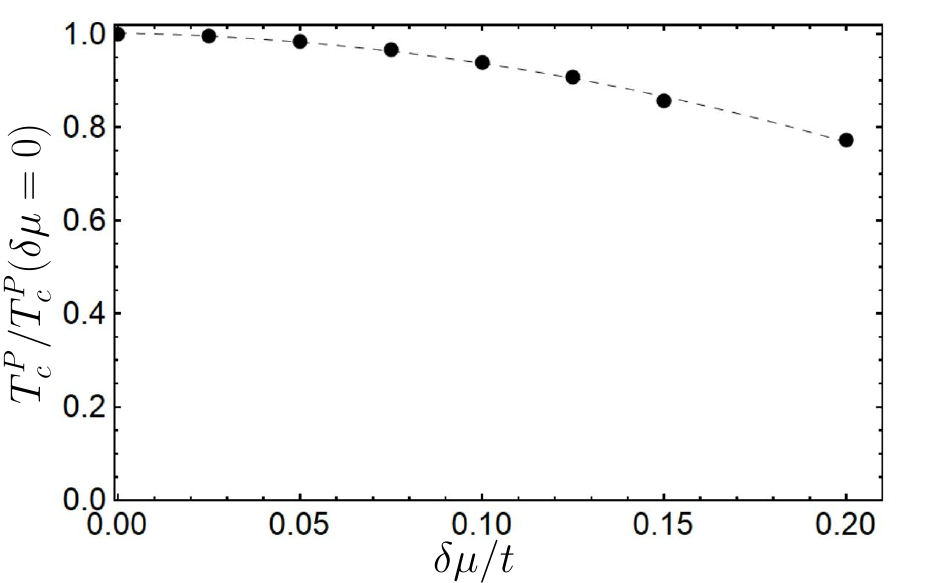}
		\caption{General suppression of $T_c$ due to impurities. $T_c(\phi=0)$ is plotted as a function of the impurity strength $\delta \mu$. At the other values of misorientation angle the critical temperature demonstrates the same general trend. $T_c$ is normalized to its value at $\delta \mu = 0$. Each point represents the averaging of the data obtained for 5-10 realizations of the random disorder restricted by a given $\delta \mu$. The dashed line is just a fit to provide a guide to eye. $\mu_S=0.9t$, $\mu_{AF}=0$, $h=t$, $d_{AF}=4$, $d_S=20$, $T_{c0}=0.03t$.}
         \label{fig:Tcphiimpurities1}
	\end{center}
\end{figure}
\begin{figure}[tb]
	\begin{center}
		\includegraphics[width=85mm]{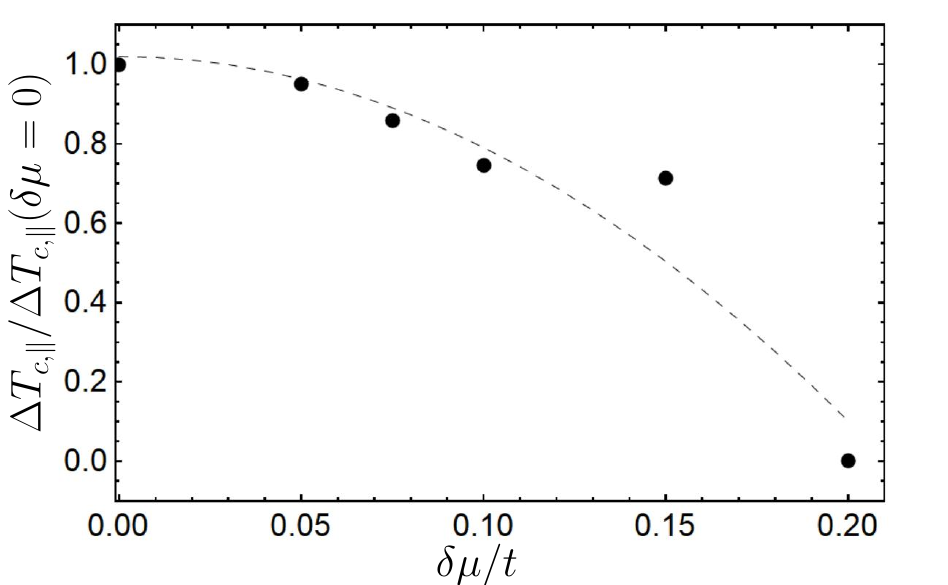}
		\caption{Suppression of the spin-valve effect by impurities. The difference $\Delta T_{c,\parallel}$ is plotted as a function of the impurity strength $\delta \mu$. The difference is normalized to its value at $\delta \mu = 0$. All parameters are the same as in Fig.~\ref{fig:Tcphiimpurities1}.}
         \label{fig:Tcphiimpurities2}
	\end{center}
\end{figure}
\begin{figure}[tb]
	\begin{center}
		\includegraphics[width=85mm]{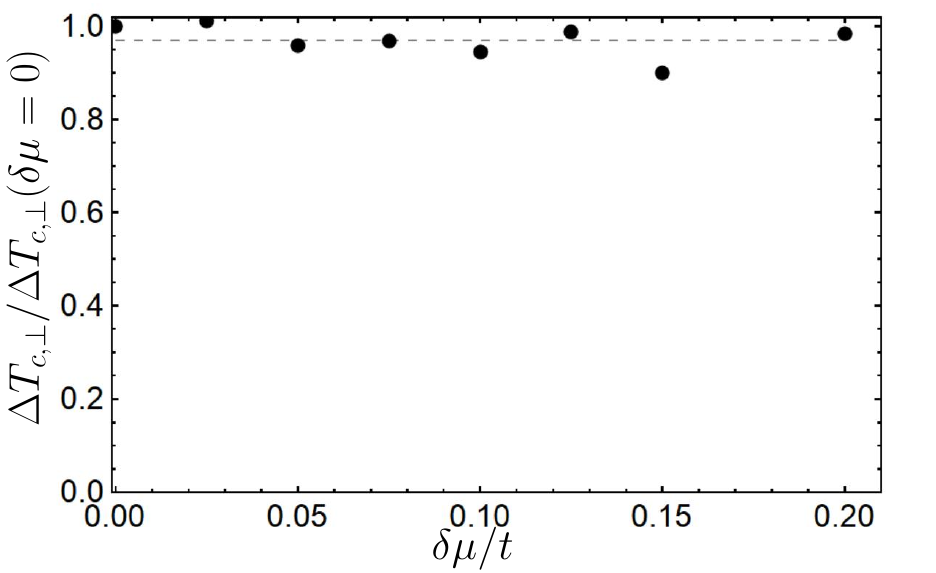}
		\caption{$\Delta T_{c,\perp}$, normalized to its value at $\delta \mu = 0$, plotted as a function of the impurity strength $\delta \mu$. All parameters are the same as in Fig.~\ref{fig:Tcphiimpurities1}.}
         \label{fig:Tcphiimpurities3}
	\end{center}
\end{figure}

In this subsection we present the results of Bogoliubov – de Gennes calculations of the critical temperature in the presence of impurities in the S region. Figures \ref{fig:Tcphiimpurities1},~\ref{fig:Tcphiimpurities2}, and \ref{fig:Tcphiimpurities3} show three different effects that we observed. All the results were obtained for the system with length $L=100$ atomic layers.

The first effect, presented in Fig.~\ref{fig:Tcphiimpurities1}, is the general suppression of the critical temperature with increasing of the impurity strength, reported in \cite{Buzdin1986,Fyhn2022,Bobkov2023}. In this subsection we set $\mu_S = 0.9t \gg T_{c0}$. In this regime, when the chemical potential is large with respect to the superconducting energy scales, the nonmagnetic impurities in the superconductor work as effectively magnetic \cite{Bobkov2023}. This results from the two sublattices and the consequent emergence of two electronic bands in the system thereby leading to the observed suppression.

Figure \ref{fig:Tcphiimpurities2} demonstrates the gradual disappearing of the valve effect under the influence of impurities, which is equivalent to the decreasing value of the difference $\Delta T_{c,\parallel} = [T_c(\phi=0)-T_c(\phi=\pi)]/2$. This is explained by the fact that the spin-valve effect is produced by the N\'eel triplets, which appear due to interband electron pairing \cite{Bobkov2022} and therefore are suppressed by impurities.

The third effect we studied is the dependence of the depth of the dips at $T_c(\phi)$ curves at $\phi=\pi/2$ on the impurity strength, which we perform by plotting the expression $\Delta T_{c,\perp} = T_c(\phi=\pi/2)-[T_c(\phi=0)+T_c(\phi=\pi)]/2$ as a function of $\delta \mu$. The results are presented in Fig.~\ref{fig:Tcphiimpurities3}. We see that this quantity tends to be insensitive to the presence of impurities. This trend is in agreement with the physical understanding that the dip is mainly produced by the cross product correlations $ f_{\mathrm{cross}}$, which are conventional (not N\'eel) triplets and correspond to intraband $s$-wave odd-frequency triplet electron pairing. Such an intraband or zero-momentum $s$-wave pairing is not suppressed by nonmagnetic impurities.

\subsection{Materials}
\label{sec:materials}

In our analysis above, we have obtained a broad picture of the spin valve effect in the trilayer system under investigation including its dependence on the chemical potential and impurity concentration. This, in turn, provides guidance with respect to the choice of materials. The predicted spin valve effect is the largest for superconductors with small chemical potential, as defined in our considerations above, which will harbor strong N\'eel correlations. Nevertheless, as it is demonstrated in the present work, the spin-valve effect is also pronounced for superconductors with large chemical potential provided that the exchange field $h$ of the AFs is strong enough. Furthermore, disorder due to impurities or interfacial lattice mismatch should be minimized. On the other hand, if one is interested in the equal-spin triplets that lead to a dip in the critical temperature around $\phi = \pi/2$, the requirements are less stringent. This is because the chemical potential can be relatively large and disorder does not play a significant role. Finally, while our theoretical considerations have focused on antiferromagnetic insulators for simplicity, one can also use metallic AFs. Having these constraints in mind, we note that our investigation is partly inspired by previous experiments observing critical temperature modifications in FeMn/Nb~\cite{Bell2003}, IrMn/Nb~\cite{Wu2013}, and IrMn/NbN~\cite{Seeger2021} bilayers. Since these combinations produce an observable effect, they should also be good candidates for investigating our proposed trilayers.

\section{Conclusions}
\label{concl}
In the present work we study AF/S/AF heterostructures with insulating antiferromagnets and fully compensated S/AF interfaces numerically by solving Bogoliubov – de Gennes equations and analytically in the framework of the  quasiclassical Green’s functions approach. We have demonstrated that the N\'eel triplet correlations lead to the dependence of superconducting critical temperature on the angle between the Néel vectors (spin-valve effect) and, in particular, to the complete suppression of superconductivity for a range of misorientation angles (absolute spin-valve effect), which is in agreement with the results presented in \cite{Johnsen_Kamra_2023}. Our calculations confirm the previous finding of \cite{Johnsen_Kamra_2023} that near half-filling the critical temperature is always lower for the parallel configuration of the N\'eel vectors, than in the antiparallel, $T_c^P<T_c^{AP}$, keeping in mind the new definitions of parallel and antiparallel used in the present manuscript.   It is explained by the fact that in this case the N\'eel triplets generated by the both interfaces are effectively summed up and strengthen each other inside the S layer. However, in the present work we investigated the spin-valve effect in the full range of $\mu_S$ values and found that if we move away from half-filling the opposite result $T_c^{P}>T_c^{AP}$ can be realized depending on the width of the S layer. This behavior results from the interference of  finite-momentum N\'eel triplet Copper pair wave functions generated by the S/AF interfaces. 

Further we investigate cross product equal-spin triplet correlations $f_{\mathrm{cross}}$, which appear in the AF/S/AF structure for non-aligned N\'eel vectors of the AFs. We provide their analytical description, prove that these correlations are of conventional (not N\'eel) $s$-wave odd-frequency type  and find parameter regions, where they are rather strong and result in the nonmonotonic dependence $T_c(\phi)$.  Finally, it is shown that the presence of impurities leads to disappearing of the ``$0-\pi$" spin-valve effect $\Delta T_{c,\parallel}=(T_c^{P}-T_c^{AP})/2$ due to the fact that impurities suppress N\'eel triplets.  At the same time, the  ``perpendicular" spin-valve effect $\Delta T_{c,\perp} = T_c(\phi=\pi/2)-(T_c^P+T_c^{AP})/2$ is not suppressed by impurities, what can be considered as a proof of its origin from  equal-spin cross product triplet correlations, which should be insensitive to impurities according to their physical nature. 

Therefore, our results significantly expand the current understanding of the physical processes in AF/S/AF spin valves, as well as they hopefully might inspire some new research in the area of spintronic devices based on proximity effects in superconductor/antiferromagnet hybrids.

\begin{acknowledgments}
The BdG analysis was supported by the Russian
Science Foundation via the RSF project No.22-22-00522. The calculations in the framework of the quasiclassical theory were supported by MIPT via Project FSMG-2023-0014. 

L.J.K., S.C., and A.K. acknowledge financial support from the Spanish Ministry for Science and Innovation -- AEI Grant CEX2018-000805-M (through the ``Maria de Maeztu'' Programme for Units of Excellence in R\&D) and grant RYC2021-031063-I funded by MCIN/AEI/10.13039/501100011033 and ``European Union Next Generation EU/PRTR''.
\end{acknowledgments}

\section*{Appendix: Calculation of the Green's function}
Let us obtain the boundary condition for the normal state Green's function at the right S/AF interface $x=d_S/2$, which is the relation between the incident and reflected Green's functions $\check g^{N}_+ (x)$ and $ \check g^{N}_- (x)$. Near the right edge of the superconductor the Eilenberger equation (\ref{Eilen}) takes the form 
\begin{align}
    \Bigl[ \Bigl(i\omega_m \tau_z+\mu_S-\bm h_{r}\bm\sigma\rho_z \tau_za\delta(x-d_S/2)\Bigr)\rho_x, \check g^{N}\Bigr]+i v_{F,x} \dfrac{d}{dx}\check g^{N}=0, 
\end{align}
which gives us the relation
\begin{align}
    i v\Bigl(\check g^{N}_- (x=d_S/2)- \check g^{N}_+ (x=d_S/2)\Bigr)=\int\limits_{C}[\bm h_r\bm\sigma a\rho_z\rho_x\tau_z\delta(x-d_S/2), \check g^{N}]dx,
    \label{integral}
\end{align}
where the integral is taken over the set $C=[d_S/2-\varepsilon, d_S/2]\cup [d_S/2, d_S/2-\varepsilon], \varepsilon \to 0. $ This leads to the boundary condition (\ref{boundary_g}). The boundary condition (\ref{boundary_f}) is obtained likewise from the Eilenberger equation (\ref{Eilen_f}) for the anomalous Green's function $\hat f$.

For simplifying the following calculations we analyze consequences of the symmetry of the considered system. Let us make the rotation by angle $\pi$ around the axis $z$:
\begin{align}
    x\to-x,~~~y\to-y,~~~z\to z.
    \label{rotation}
\end{align}
After rotation (\ref{rotation}) different objects which are present in our problem are transformed in the following way: scalars in spin space do not change ($A(x)\to A(-x)$), components of vectors in spin space change as $A_x(x)\to-A_x(-x), A_y(x)\to -A_y(-x), A_z(x)\to A_z(-x)$. Due to the symmetric choice of the coordinate axes the system itself does not change after rotation (\ref{rotation}), but the reflected Green's function goes to the incident one. This gives us the following relations:
\begin{align}
\begin{cases}
    \check g_{0\alpha,-}(x)=\check g_{0\alpha,+}(-x)\\
    \check g_{x\alpha,-}(x)=-\check g_{x\alpha,+}(-x)\\
    \check g_{y\alpha,-}(x)=-\check g_{y\alpha,+}(-x)\\
    \check g_{z\alpha,-}(x)=\check g_{z\alpha,+}(-x),
    \end{cases}
    \label{symm}
\end{align}
where 
\begin{align}
\check g_{\beta\alpha, +(-)} \equiv
\left(
\begin{array}{cc}
g_{\beta\alpha, +(-)} & f_{\beta\alpha, +(-)} \\
{\tilde f}_{\beta\alpha, +(-)} & {\tilde g}_{\beta\alpha, +(-)}
\end{array}
\right).   
\end{align}
In the following text Green's functions without indices $+(-)$ correspond to incident trajectories, and the reflected Green's functions can be obtained from the relations (\ref{symm}). 

Now let us note that the simultaneous transformation from sublattice $A$ to sublattice $B$ and $\bm h \to -\bm h$ does not change the system and, therefore, does not change the Green's function. Therefore, if we write $\check g$ in the form $\check g=\check g^0 + \check g^h$, where $\check g^0$ and $\check g^h$ are even and odd functions with respect to $h \to -h$, the transformation $A\leftrightarrow B$ leads to $\check g^0\to \check g^0$ and $\check g^h\to -\check g^h$. From the other hand, the transformation $A \leftrightarrow B$ changes sign of $\rho_{y,z}$-components of the Green's function in the sublattice space, remaining $\rho_{0,x}$-components unchanged. Consequently, $\check g^0$ has non-zero $\rho_{0,x}$-components, and  $\check g^h$ has non-zero $\rho_{y,z}$-components. Then, from two vectors $\bm h_{l,r}$, which characterize our system, we can make the following basic combinations, which describe the Green's function in the spin space: $\bm h_l$, $\bm h_r$, $\bm h_l \bm h_r$ and $\bm h_l \times \bm h_r$. The first two items are odd functions with respect to $h \to -h$ and enter into the $\sigma_{y,z}$-components of the Green's function, and the second two items are even functions and enter  into the $\sigma_{0,x}$-components of the Green's function.  Thus, we can conclude that $\check g_0$ and $\check g_x$ have only $\sigma_0,\sigma_x$ non-zero components in the expansion over Pauli matrices in the spin space, while $\check g_y$ and $\check g_z$ have only $\sigma_y,\sigma_z$ non-zero components. Another consequence of the symmetry towards rotation (\ref{rotation}) and the structure of boundary conditions (\ref{boundary_g}), (\ref{boundary_f}) is the following expanding of (\ref{symm}), which can also be considered an ansatz:

\begin{align}
    \begin{cases}
    \check g_{zy,+}(x)=\check g_{zy,+}(-x)=\check g_{zy,-}(x)\\
    \check g_{zz,+}(x)=-\check g_{zz,+}(-x)=-\check g_{zz,-}(x)\\
    \check g_{yy,+}(x)=-\check g_{yy,+}(-x)=\check g_{yy,-}(x)\\
    \check g_{yz,+}(x)=\check g_{yz,+}(-x)=-\check g_{yz,-}(x).
    \end{cases}
    \label{sym_final}
\end{align}

The Eilenberger equation on the particle component $\hat g$ of the normal state Green's function takes the form
\begin{align}
    [(i\omega_m+\mu_S)\rho_x, \hat g]+iv_{F,x}\dfrac{d}{dx}\hat g=0
\end{align}
apart from the S/AF interfaces and can be expanded over components in the sublattice space:
\begin{align}
    \begin{cases}
    \dfrac{d}{dx} \hat g_0=0 \\
    \dfrac{d}{dx} \hat g_x=0 \\
    2(-\omega_m+i\mu_S)\hat g_y+iv_{F,x}\dfrac{d}{dx} \hat g_z=0 \\
    2(\omega_m -i\mu_S)\hat g_z+iv_{F,x}\dfrac{d}{dx} \hat g_y=0.
    \end{cases}
\end{align}
As was discussed above, the components $g_{xy},g_{xz},g_{0y},g_{0z},g_{z0},g_{zx},g_{y0},g_{yx}$ are equal to zero because of symmetry. Considering (\ref{sym_final}), we write the solutions for other components:
\begin{align}
    \begin{cases}
    g_{00}=const \\
    g_{x0}=const \\
    g_{0x}=const \\
    g_{xx}=const \\
    g_{yy}=i B \sinh \kappa x \\
    g_{zy}=-A \cosh \kappa x \\
    g_{yz}=B ~\textrm{sgn}~ v_{F,x} \cosh \kappa x\\
    g_{zz}=i A ~\textrm{sgn}~ v_{F,x} \sinh \kappa x,
    \end{cases}
    \label{g_sol}
\end{align}
where $\kappa=2(i\mu_S-\omega_m)/v$ and $A, B$ are unknown coefficients. From the boundary condition (\ref{boundary_g}) we obtain  
\begin{align}
    \begin{cases}
    \delta g_{00}=0 \\
    \delta g_{x0}=-\Bigl(g_{yy}(x=d_S/2)\gamma_z-g_{zy}(x=d_S/2)\gamma_y\Bigr) \\
    \delta g_{0x}=+\Bigl(g_{zz}(x=d_S/2)\gamma_z+g_{yz}(x=d_S/2)\gamma_y\Bigr) \\
    \delta g_{xx}=0 \\
    \delta g_{yy}=+g_{x0}(x=d_S/2)\gamma_z \\
    \delta g_{zy}=-g_{x0}(x=d_S/2)\gamma_y \\
    \delta g_{yz}=-g_{0x}(x=d_S/2)\gamma_y \\
    \delta g_{zz}=-g_{0x}(x=d_S/2)\gamma_z,
    \end{cases}
    \label{bound_g}
\end{align}
where $\gamma_{y,z}=4i a h_{y,z}/v$. Eq.~(\ref{bound_g}) gives us $g_{00}=g_{xx}=0$. Coefficients $g_{x0}, g_{0x}, A$ and $B$ can be found from the normalization condition (\ref{norm}), which takes the form 
\begin{align}
   A^2+B^2+g_{0x}^2+g_{x0}^2=1,
\end{align}
and substituting solutions (\ref{g_sol}) for $g_{yy}, g_{yz}, g_{zy}, g_{zz}$ into the last 4 equations of the system (\ref{bound_g}). We obtain:
\begin{align}
    g_{0x}=\dfrac{1}{\sqrt{(1+\gamma_a^2)(1+\gamma_b^2)}},\nonumber \\    g_{x0}=\dfrac{\gamma_a\gamma_b}{\sqrt{(1+\gamma_a^2)(1+\gamma_b^2)}},\nonumber \\
    A=\dfrac{\gamma_a}{\sqrt{(1+\gamma_a^2)(1+\gamma_b^2)}},\nonumber \\
    B=\dfrac{\gamma_b}{\sqrt{(1+\gamma_a^2)(1+\gamma_b^2)}},
\end{align}
where $\gamma_a=-i\gamma_z/2\sinh (\kappa d_S/2), \gamma_b=\gamma_y/2\cosh (\kappa d_S/2).$ In the first order with respect to $\kappa d_S$ $\gamma_a=2a h_z/(-\omega_m+i\mu_S)d_S, \gamma_b=-2a h_y/v.$

The Eilenberger equation on the anomalous Green's function $\hat f$ takes the form 
\begin{align}
     i\omega_m\{\rho_x, \hat f\}+\mu_S [\rho_x, \hat f]+\Delta(\rho_x\hat {\tilde g}-\hat g\rho_x)+iv_{F,x}\dfrac{d}{dx}\hat f=0
\end{align}
apart from the interfaces and can be expanded over components in the sublattice space:
\begin{align}
    \begin{cases}
        2i\omega_m \hat f_x+iv_{F,x}\dfrac{d}{dx} \hat f_0=\Delta(\hat g_x-\hat {\tilde g}_x)\equiv\hat G_x \\
        2i\omega_m \hat f_0+iv_{F,x}\dfrac{d}{dx} \hat f_x=\Delta(\hat g_0-\hat {\tilde g}_0)\equiv\hat G_0 \\
        -2i\mu_S \hat f_z+iv_{F,x}\dfrac{d}{dx}\hat f_y=i\Delta(\hat g_z+\hat {\tilde g}_z)\equiv\hat G_z \\
        2i\mu_S \hat f_y+iv_{F,x}\dfrac{d}{dx} \hat f_z=-i\Delta(\hat g_y+\hat {\tilde g}_y)\equiv\hat G_y.
    \end{cases}
    \label{system_f}
\end{align} 
Let us expand $\hat G_{y,z}$ up to the first order with respect to $x$: $\hat G_{y,z}=\hat G_{y,z}^0+\hat G'_{y,z}x$. Then the general solution of (\ref{system_f}) is
\begin{align}
    \begin{cases}
        \hat f_0=\hat A\sinh\Bigl(\dfrac{2\omega_m}{v}x\Bigr)+\hat B\cosh\Bigl(\dfrac{2\omega_m}{v}x\Bigr)+\dfrac{\hat G_0}{2i\omega_m}\\
        \hat f_x=-\hat A\cosh\Bigl(\dfrac{2\omega_m}{v}x\Bigr)-\hat B\sinh\Bigl(\dfrac{2\omega_m}{v}x\Bigr)+\dfrac{\hat G_x}{2i\omega_m}\\
        \hat f_y=\hat C\sin\Bigl(\dfrac{2\mu_S}{v}x\Bigr)+\hat D\cos\Bigl(\dfrac{2\mu_S}{v}x\Bigr)+\dfrac{\hat G_y^0+\hat G'_y x}{2i\mu_S}+\dfrac{v}{4i\mu_S^2}\hat G'_z\\
        \hat f_z=\hat C\cos\Bigl(\dfrac{2\mu_S}{v}x\Bigr)-\hat D\sin\Bigl(\dfrac{2\mu_S}{v}x\Bigr)-\dfrac{\hat G_z^0+\hat G'_z x}{2i\mu_S}+\dfrac{v}{4i\mu_S^2}\hat G'_y,      
    \end{cases}
\end{align}
where $\hat A, \hat B, \hat C, \hat D$ are $2\times2$ spin matrices of unknown coefficients. $f_{0y},f_{0z},f_{xy},f_{xz},f_{y0},f_{yx},f_{z0},f_{zx}$ are equal to zero due to symmetry, and for other components we obtain in the linear order with respect to $x$:
\begin{align}
    \begin{cases}
        f_{00}=A_0\dfrac{2\omega_m}{v}x~\textrm{sgn}~v_{F,x}+B_0+\dfrac{G_{00}}{2i\omega_m} \\
        f_{0x}=-A_0-B_0\dfrac{2\omega_m}{v}x~\textrm{sgn}~v_{F,x}+\dfrac{G_{0x}}{2i\omega_m} \\
        f_{x0}=A_x\dfrac{2\omega_m}{v}x+\Bigl(B_x+\dfrac{G_{x0}}{2i\omega_m}\Bigr)~\textrm{sgn}~v_{F,x} \\
        f_{xx}=\Bigl(-A_x+\dfrac{G_{xx}}{2i\omega_m}\Bigr)~\textrm{sgn}~v_{F,x}-B_x\dfrac{2\omega_m}{v}x \\
        f_{yy}=\Bigl(C_y\dfrac{2\mu_S}{v}+\dfrac{G'_{yy}}{2i\mu_S}\Bigr)x+D_y~\textrm{sgn}~v_{F,x}+\Bigl(G_{yy}^0+\dfrac{v}{2\mu_S}G'_{yz}\Bigr)\dfrac{\textrm{sgn}~v_{F,x}}{2i\mu_S} \\
        f_{yz}=C_y~\textrm{sgn}~v_{F,x}-\Bigl(D_y\dfrac{2\mu_S}{v}+\dfrac{G'_{yz}}{2i\mu_S}\Bigr)x+\Bigl(-G_{yz}^0+\dfrac{v}{2\mu_S}G'_{yy}\Bigr)\dfrac{\textrm{sgn}~v_{F,x}}{2i\mu_S}\\
        f_{zy}=\Bigl(C_z\dfrac{2\mu_S}{v}+\dfrac{G'_{zy}}{2i\mu_S}\Bigr)x~\textrm{sgn}~v_{F,x}+D_z+\Bigl(G_{zy}^0+\dfrac{v}{2\mu_S}G'_{zz}\Bigr)\dfrac{1}{2i\mu_S}\\
        f_{zz}=C_z-\Bigl(D_z\dfrac{2\mu_S}{v}+\dfrac{G'_{zz}}{2i\mu_S}\Bigr)x~\textrm{sgn}~v_{F,x}+\Bigl(-G_{zz}^0+\dfrac{v}{2\mu_S}G'_{zy}\Bigr)\dfrac{1}{2i\mu_S},
    \end{cases}
\end{align}
where we have used the expansions $\hat A=\sum\limits_\alpha\sigma_\alpha A_\alpha, \hat B=\sum\limits_\alpha\sigma_\alpha B_\alpha, \hat C=\sum\limits_\alpha\sigma_\alpha C_\alpha, \hat D=\sum\limits_\alpha\sigma_\alpha D_\alpha, \hat G_\beta=\sum\limits_\alpha\sigma_\alpha G_{\alpha\beta}$. The solution for the normal Green's function gives us $G'_{yz}=G'_{zy}=G_{yy}^0=G_{zz}^0=G_{00}=G_{xx}=0$, and the relations (\ref{sym_final}) lead to $B_0=A_x=D_y=C_z=0$. Other unknown coefficients are found from the boundary condition (\ref{boundary_f}):
\begin{align}
\begin{cases}
    A_0=\dfrac{i(G'_{zz} h_z a v^4+4\mu_S(v^2(G_{zy}^0 h_z a v+d_S G_{yz}^0 h_y a\mu_S)-2G_{x0}h_yh_z a^2(v^2+d_S^2\mu_S^2)))}{4 d_S v^3 \mu_S^2 \omega_m} \\ 
    B_x=\dfrac{i G_{x0}}{2\omega_m} \\ 
    C_y=\dfrac{i (G'_{yy}v^3+2\mu_S(-G_{yz}^0 v^2+2d_S G_{x0}h_z a\mu_S))}{4v^2\mu_S^2} \\ 
    D_z=\dfrac{i(G'_{zz} v^2+8G_{x0} h_y a \mu_S)}{8v\mu_S^2}.
    \end{cases}
\end{align}

\bibliography{spin_valves}

\end{document}